\documentclass[showpacs,10pt,twocolumn,prb]{revtex4-1}
\usepackage{amsmath}
\usepackage{amssymb}
\usepackage{graphics}
\usepackage{epsfig}

\begin{document}

\title{Magnetism in La$_{2}$O$_{3}$(Fe$_{1-x}$Mn$_{x}$)$_{2}$Se$_{2}$ tuned
by Fe/Mn ratio}
\author{Hechang Lei,$^{1}$ Emil S. Bozin,$^{1}$ A. Llobet,$^{2}$ V.
Ivanovski,$^{3}$ V. Koteski,$^{3}$ J. Belosevic-Cavor,$^{3}$ B. Cekic,$^{3}$
and C. Petrovic$^{1}$}
\affiliation{$^{1}$Condensed Matter Physics and Materials Science Department, Brookhaven National Laboratory, Uptown, NY 11973 USA\\
$^{2}$Lujan Neutron Scattering Center, Los Alamos National Laboratory, MS H805, Los Alamos, New Mexico 87545, USA\\
$^{3}$Institute of Nuclear Sciences Vinca, University of Belgrade, Belgrade 11001, Serbia}
\date{\today}

\begin{abstract}
We report the evolution of structural and magnetic properties in La$_{2}$O$%
_{3}$(Fe$_{1-x}$Mn$_{x}$)$_{2}$Se$_{2}$. Heat capacity and bulk
magnetization indicate increased ferromagnetic component of the long range
magnetic order and possible increased degree of frustration. Atomic disorder
on Fe(Mn) sites suppresses the temperature of the long range order whereas
intermediate alloys show rich magnetic phase diagram.
\end{abstract}


\maketitle

\section{Introduction}

Parent materials of cuprates and iron-based superconductors are layered
transition-metal compounds (TMP's) with antiferromagnetic (AFM) ground state
due to a dissimilar magnetic mechanism and with rather different electronic
conductivity.\cite{Mazin} Local environment of TM is intimately connected
with this since puckering of iron pnictide crystallographic layers promotes
higher conductivity in parent materials of iron superconductors, as opposed
to cuprates.\cite{Cvetkovic,Qazilbash} Recently, a family of layered TMP
oxychalcogenides Ln$_{2}$O$_{3}$TM$_{2}$Ch$_{2}$ (Ln = rare earth, TM =
transition metal, and Ch = S, Se) and its analogues A$_{2}$F$_{2}$TM$_{2}$OCh%
$_{2}$ (A = Sr, Ba)(2322) has received more attention.\cite%
{Mayer,Clarke,Kabbour,Wang,Zhu,Free1,Wu,Fuwa1,Fuwa2,McCabe,Ni,Liu1,Free2}
They have anti-CuO$_{2}$-type [TM$_{2}$OCh$_{2} $]$^{2-}$ layers and the
local environment of TM bears much more resemblance to manganites or
cuprates than to iron based superconductors. This typically results in
semiconducting or Mott insulating ground state with AFM transition.\cite%
{Clarke,Zhu} Within the [TM$_{2}$OCh$_{2} $]$^{2-}$ layers transition metal
atoms are located at the distorted octahedral environment due to the
different distances of TM-Ch and TM-O, whereas the TM-Ch(O) octahedrons are
face sharing. This distortion lifts the degeneracy of $e_{g}$ and $t_{2g}$
levels in a non-distorted octahedral crystal field. The [TM$_{2}$OCh$_{2} $]$%
^{2-}$ layers in the crystal structure harbor three principal competing
interactions: the nearest neighbor (NN) $J_{3}$ interaction of the closest
TM atoms in the crystallographic $ab$ plane, next nearest neighbor (NNN)
superexchange interaction $J_{2}$ of nearly 90$^{\circ }$ TM-Ch-TM, and
(NNN) $J_{1}$ superexchange interaction of 180$^{\circ }$ TM-O-TM.\cite%
{Mayer,Kabbour,Zhu,Ni} Theoretical calculation and neutron diffraction
results indicate that the AFM transition stems from the competition of three
interactions in an unusual frustrated AFM checkerboard spin-lattice system.%
\cite{Kabbour,Wang,Zhu,Free1,Wu,Ni,Free2} Frustrated magnetic order on
checkerboard lattices hosts rich magnetic ground states and could be
relevant to physics of colossal magnetoresistance manganites and iron based
and cuprates superconducting materials.\cite{Ling,Low,Yildirim,Si,Han}
Therefore it is of interest to address the nature of magnetic order and
frustration in 2322 materials since it shares some characteristics of all
the above mentioned compounds.

Among 2322 materials, La$_{2}$O$_{3}$Mn$_{2}$Se$_{2}$ (Mn-2322) exhibits a
faint specific heat anomaly at AFM transition and two dimensional (2D) short
range magnetic correlations that lock magnetic entropy far above the
temperature of the bulk long range AFM transition.\cite{Ni,Liu1} Weak
ferromagnetic (FM) component superimposed on the paramagnetic background in
M(H) loops at 2 K proposed to arise due to spin reorientation and/or spin
canting below the AFM transition,\cite{Ni} might also come from magnetic
impurities such as Mn$_{3}$O$_{4}$.\cite{Free2} Mn-2322 has a G-type AFM
structure with the ordered moment along the $c$ axis direction and where NN
Mn ions have opposite spins with dominant AFM $J_{3}<0$ and fully frustrated
$|J_{1}|<J_{2}$ with $J_{1}<0$ and $J_{2}>0$.\cite{Ni,Free2} On the other
hand, La$_{2}$O$_{3}$Fe$_{2}$Se$_{2}$ (Fe-2322) adopts magnetic structure
similar to FeTe where half of the $J_{3}$, $J_{2}$ and $J_{1}$ are frustrated%
\cite{Free1,Free2,Li} and different from the theoretical predictions.\cite%
{Zhu} This is different from other compounds with same structure, in which $%
J_{1}$ is the dominant term.\cite{Wu}

In this work we tuned the magnetism on Fe/Mn checkerboard lattice by varying
TM content and examined the magnetic ground states of La$_{2}$O$_{3}$(Fe$%
_{1-x}$Mn$_{x}$)$_{2}$Se$_{2}$ series. We find that Mn substitution lead to
the increase of activation energies for the electrical transport energy gap $%
E_{a}$ and has strong effect on magnetic phase diagram and the spin entropy
release above the temperature of the long range magnetic order.

\section{Experiment}

La$_{2}$O$_{3}$(Fe,Mn)$_{2}$Se$_{2}$ ((Fe,Mn)-2322) polycrystals were
synthesized by solid state reaction from high purity materials ($\geqslant $
99\%). Dried La$_{2}$O$_{3}$ powder, Mn lump, Fe and Se powders were mixed
and ground in an agate mortar. The mixture was pressed into pellets and
sealed in a quartz tube backfilled with pure Argon gas. The ampule was
heated to 1273 K and reacted for 24 h followed by furnace cooling. This was
repeated several times to ensure homogeneity. The color of final product is
yellowish green for Mn-2223 and black for Fe-2322.

Powder X-ray diffraction (XRD) patterns of the ground samples were taken
with Cu $K_{\alpha }$ = 0.1458 nm using a Rigaku Miniflex X-ray machine. The
structural parameters were obtained by Rietveld refinement using RIETICA
software.\cite{Hunter} Single crystal X-ray data were collected in a Bruker
Kappa Apex II single crystal X-ray diffractometer at the room temperature
with Mo $K_{\alpha }$ = 0.071073 nm on a 20 $\mu $m single crystal isolated
from the polycrystalline powder. Single crystal unit cell refinement was
performed with Bruker APEX 2 software package. Time-of-flight neutron
diffraction measurements were carried out at 300 K on the high intensity
powder diffractometer (HIPD) of the Lujan Neutron Scattering Center at the
Los Alamos National Laboratory. Pulverized samples were loaded in extruded
vanadium containers in helium atmosphere and sealed. Data were collected for
1 hour for each sample. Rietveld refinements of the data were carried out
using GSAS software\cite{Larson,Toby} within the I4/mmm space group.

M\"{o}sbauer spectra were taken in transmission mode with $^{57}$Co(Rh)
source at 294 K and the parameters were obtained using WinNormos software.%
\cite{Brand} Calibration of the spectrum was performed by laser and isomer
shifts were given with respect to $\alpha $-Fe.

Thermal, transport and magnetic measurements were carried out in a Quantum
Design PPMS-9 and MPMS-5. The samples were cut into rectangular
parallelepipeds and thin Pt wires were attached for four probe resistivity
measurements. Sample dimensions were measured with an optical microscope
Nikon SMZ-800 with 10 $\mu $m resolution.

\section{Results and Discussions}

\begin{figure}[tbp]
\centerline{\includegraphics[scale=0.40]{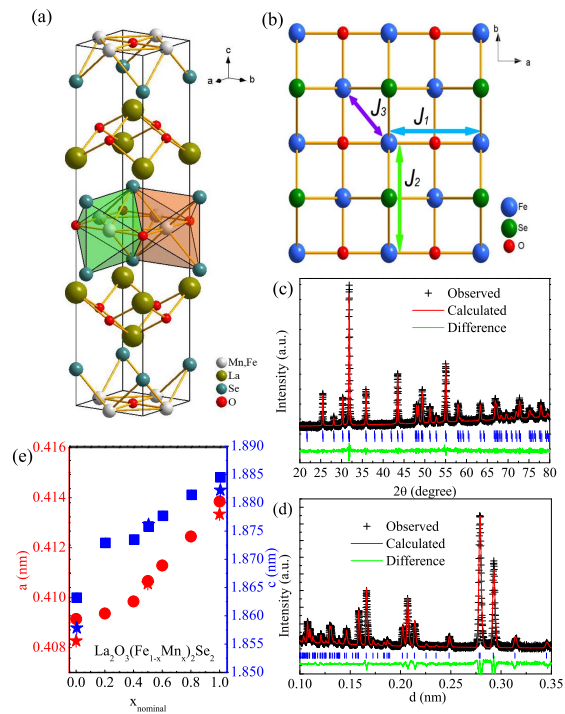}} \vspace*{-0.3cm}
\caption{(a) Crystal structure of (Fe,Mn)-2322. (b) TM-O planes shown for TM
= Fe. Se atoms (green) are puckered above and below the plane, with Se-Fe-Se
angle 97.04$^{\circ }$ along both $a$ and $b$ crystallographic axes. (c) and
(d) XRD and neutron powder diffraction results at room temperature for the
Mn-2322 polycrystal, respectively. (e) Lattice parameters for (Fe,Mn)-2322
determined from neutron diffraction (stars) and XRD (circles).}
\end{figure}

Crystal structure of (Fe,Mn)-2322 and principle magnetic interactions $J_{1}$%
-$J_{3}$ are shown in Fig. 1(a,b). Both XRD and neutron reflections can be
indexed using a tetragonal unit cell (Fig. 1(c,d)), indicating a continuous
solid solution between Mn-2322 and Fe-2322. Unit cell of Mn-2322 single
crystal determined from the single crystal X-ray experiment was I4/mmm with $%
a$ = 0.417(3) nm and $c$ = 1.89(2) nm, in agreement with the powder data.
Powder X-ray and neutron lattice parameters are shown in Fig. 1(e) and
details of the crystal structure extracted from the neutron data, listed in
Tables I-III, are consistent with the reported values.\cite{Zhu,Ni} For
Mn-2322, using obtained Mn-Se and Mn-O bond lengths (shown in Table 1), we
calculate the valence of Mn ions using the bond valence sum (BVS) formalism
in which each bond with a distance $d_{ij}$ contributes a valence $%
v_{ij}=\exp [(R_{ij}-d_{ij})/0.37]$ with $R_{ij}$ as an empirical parameter
and the total of valences of atom i, $V_{i}$ equals $V_{i}=\underset{j}{\sum
v_{ij}}$.\cite{Brown,Brese} The calculated valence of Mn ions is +2.03,
consistent with the apparent oxidation state (+2) for Mn atoms. With
increase in Mn content, the lattice parameters increase gradually for both $%
a $ and $c$ axes, hence the unit cell of Mn-2322 is larger than Fe-2322,
which itself is larger than the unit cell of La$_{2}$O$_{2}$Co$_{2}$OSe$_{2}$
(Co-2322).\cite{Mayer}$^{,}$\cite{Wang} This trend can be explained by the
larger ionic size of Mn$^{2+}$ when compared to Fe$^{2+}$ and Co$^{2+}$
ions, giving some insight in the spin state of Mn$^{2+}$ ions. The lattice
parameters should be proportional to the ionic size, $r_{TM^{2+}}$, of
transition metals TM$^{2+}$ (TM = Mn, Fe, and Co). Therefore from the
observation $a_{Mn}$ = 0.41334 nm $>$ $a_{Fe}$ = 0.40788 nm $>$ $a_{Co}$ =
0.40697 nm and $c_{Mn}$ = 1.88221 nm $>$ $c_{Fe}$ = 1.86480 nm $>$ $c_{Co}$=
1.84190 nm we conclude that $r_{Mn^{2+}}>r_{Fe^{2+}}>r_{Co^{2+}}$. M\"{o}%
sbauer spectra and theoretical calculations for Fe-2322 and Co-2322 suggest
that Fe$^{2+}$ and Co$^{2+}$ are in the high spin state.\cite{Kabbour}$^{,}$%
\cite{Wu} This is in agreement with $r_{Fe^{2+}}$(0.0780 nm) $>$ $%
r_{Co^{2+}} $(0.0745 nm) for high spin (HS) state (coordination number = 6).
Therefore, the Mn$^{2+}$ ions should be in the HS state since $r_{Mn^{2+}}$%
(LS) = 0.067 nm (LS = low spin) and $r_{Mn^{2+}}$(HS) = 0.083 nm for Mn$%
^{2+} $ with 6-fold coordination.

\begin{table}[tbp]\centering%
\caption{Structural parameters for La$_{2}$O$_{3}$Fe$_{2}$Se$_{2}$ at 300 K.}%
\begin{tabular}{cccccc}
\hline\hline
\multicolumn{3}{c}{Chemical Formula} & \multicolumn{3}{c}{La$_{2}$O$_{3}$Fe$%
_{2}$Se$_{2}$} \\ \hline
\multicolumn{3}{c}{Interatomic Distances (nm)} & \multicolumn{3}{c}{Bond
Angles ($^{\circ }$)} \\
\multicolumn{2}{c}{$d_{Fe-O}$} & 0.20413(1) & \multicolumn{2}{c}{O-Fe-O} &
180 \\
\multicolumn{2}{c}{$d_{Fe-Se}$} & 0.27129(8) & \multicolumn{2}{c}{Se-Fe-Se}
& 97.61(4)/82.39(4) \\
\multicolumn{2}{c}{$d_{Fe-Fe}$} & 0.28869(1) & \multicolumn{2}{c}{Se-Fe-O} &
90 \\ \hline
Atom & x & y & z & Occ & $U_{iso}$ (10$^{-2}$nm$^{2}$) \\
La & 0.5 & 0.5 & 0.1843(1) & 1 & 0.0026(2) \\
O1 & 0.5 & 0 & 0.25 & 1 & 0.0059(3) \\
O2 & 0.5 & 0.5 & 0 & 1 & 0.0124(5) \\
Fe & 0.5 & 0 & 0 & 1 & 0.0059(2) \\
Se & 0 & 0 & 0.0962(1) & 1 & 0.0032(2) \\ \hline\hline
\end{tabular}%
\label{1}%
\end{table}%

\begin{table}[tbp]\centering%
\caption{Structural parameters for
La$_{2}$O$_{3}$(Fe$_{0.5}$Mn$_{0.5}$)$_{2}$Se$_{2}$ at 300 K.}%
\begin{tabular}{cccccc}
\hline\hline
\multicolumn{3}{c}{Chemical Formula} & \multicolumn{3}{c}{La$_{2}$O$_{3}$(Fe$%
_{0.5}$Mn$_{0.5}$)$_{2}$Se$_{2}$} \\ \hline
\multicolumn{3}{c}{Interatomic Distances (nm)} & \multicolumn{3}{c}{Bond
Angles ($^{\circ }$)} \\
\multicolumn{2}{c}{$d_{Fe/Mn-O}$} & 0.20528(1) & \multicolumn{2}{c}{O-Fe-O}
& 180 \\
\multicolumn{2}{c}{$d_{Fe/Mn-Se}$} & 0.27641(12) & \multicolumn{2}{c}{
Se-Fe-Se} & 95.92(6)/84.08(6) \\
\multicolumn{2}{c}{$d_{Fe/Mn-Fe/Mn}$} & 0.29031(1) & \multicolumn{2}{c}{
Se-Fe/Mn-O} & 90 \\ \hline
Atom & x & y & z & Occ & $U_{iso}$ (10$^{-2}$nm$^{2}$) \\
La & 0.5 & 0.5 & 0.1851(1) & 1 & 0.0057(3) \\
O1 & 0.5 & 0 & 0.25 & 1 & 0.0051(5) \\
O2 & 0.5 & 0.5 & 0 & 1 & 0.0100(7) \\
Fe & 0.5 & 0 & 0 & 1 & 0.0058(8) \\
Se & 0 & 0 & 0.0987(1) & 1 & 0.0044(4) \\ \hline\hline
\end{tabular}%
\label{2}%
\end{table}%

\begin{table}[tbp]\centering%
\caption{Structural parameters for La$_{2}$O$_{3}$Mn$_{2}$Se$_{2}$ at 300 K.}%
\begin{tabular}{cccccc}
\hline\hline
\multicolumn{3}{c}{Chemical Formula} & \multicolumn{3}{c}{La$_{2}$O$_{3}$Mn$%
_{2}$Se$_{2}$} \\ \hline
\multicolumn{3}{c}{Interatomic Distances (nm)} & \multicolumn{3}{c}{Bond
Angles ($^{\circ }$)} \\
\multicolumn{2}{c}{$d_{Mn-O}$} & 0.2667(1) & \multicolumn{2}{c}{O-Mn-O} & 180
\\
\multicolumn{2}{c}{$d_{Mn-Se}$} & 0.28030(1) & \multicolumn{2}{c}{Se-Mn-Se}
& 95.01(1)/84.99(1) \\
\multicolumn{2}{c}{$d_{Mn-Mn}$} & 0.29227(1)) & \multicolumn{2}{c}{Se-Fe-O}
& 90 \\ \hline
Atom & x & y & z & Occ & $U_{iso}$ (10$^{-2}$nm$^{2}$) \\
La & 0.5 & 0.5 & 0.1866(2) & 1 & 0.0042(5) \\
O1 & 0.5 & 0 & 0.25 & 1 & 0.0057(7) \\
O2 & 0.5 & 0.5 & 0 & 1 & 0.0015(8) \\
Mn & 0.5 & 0 & 0 & 1 & 0.0086(8) \\
Se & 0 & 0 & 0.1006(1) & 1 & 0.0020(5) \\ \hline\hline
\end{tabular}%
\label{3}%
\end{table}%

\begin{table}[tbp]\centering%
\caption{M\"{o}sbauer isomer shifts $\delta$ and electric field quadrupole
splittings $\Delta$E$_{Q}$ for
La$_{2}$O$_{3}$(Fe$_{1-x}$Mn$_{x}$)$_{2}$Se$_{2}$ at 294 K.}%
\begin{tabular}{cccccc}
\hline\hline\hline
Chemical formula & $\delta $(mm/s) & $\Delta $E$_{Q}$(mm/s) & $\Gamma $(mm/s)
&  &  \\
La$_{2}$O$_{3}$Fe$_{2}$Se$_{2}$ & 0.895(1) & 1.952(1) & 0.287(2) &  &  \\
La$_{2}$O$_{3}$(Fe$_{0.8}$Mn$_{0.2}$)$_{2}$Se$_{2}$ & 0.90(4) & 1.756 & 0.309
&  &  \\
La$_{2}$O$_{3}$(Fe$_{0.5}$Mn$_{0.5}$)$_{2}$Se$_{2}$ & 0.812(6) & 1.60(1) &
0.309(7) &  &  \\
&  &  &  &  &  \\ \hline\hline
\end{tabular}%
\label{4}%
\end{table}%

\begin{figure}[tbp]
\centerline{\includegraphics[scale=0.4]{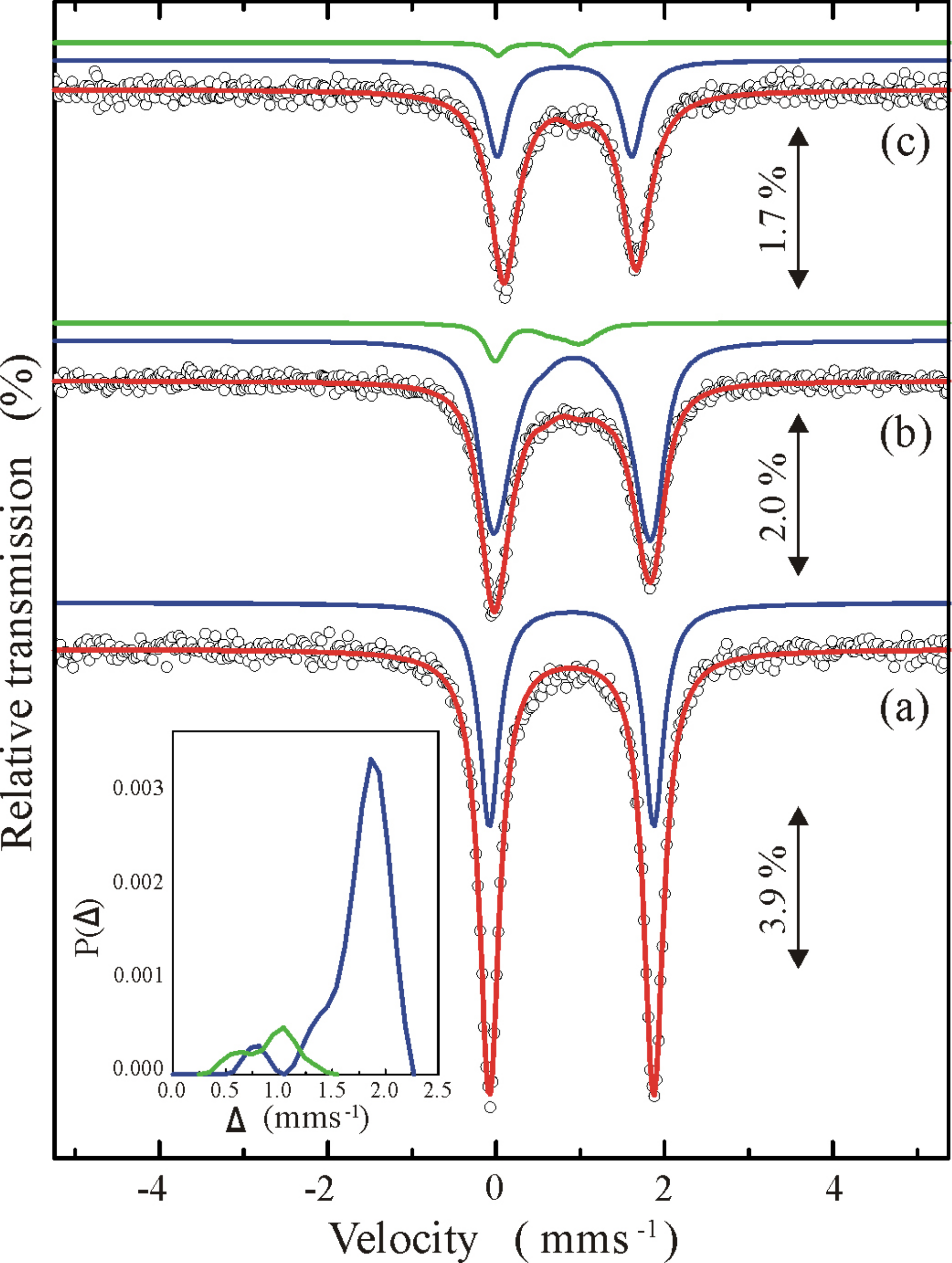}} \vspace*{-0.3cm}
\caption{M\"{o}sbauer spectra of La$_{2}$O$_{3}$(Fe$_{1-x}$Mn$_{x}$)$_{2}$Se$%
_{2}$ for $x$ = 0 (a), $x$ = 0.2 (b), and $x$ = 0.5 (c), respectively.
Vertical arrows denote relative positions of peaks with respect to the
background. The data were shown by open circles and the fit is given by the
red solid line. The doublets of the main (blue) and impurity (green) phases
are offset for clarity. Inset shows the distributions of the electrical
quadrupole splitting at the $^{57}$Fe spectrum of La$_{2}$O$_{3}$(Fe$_{0.8}$%
Mn$_{0.2}$)$_{2}$Se$_{2}$.}
\end{figure}

M\"{o}sbauer isomer shifts $\delta $ are sensitive on the electron density
on a nucleus. Electric quadrupole splitting $\Delta $ arises due to
interaction between electric field gradient and nuclear quadrupole moment.
The $\delta $ and $\Delta $ provide information of the oxidation, spin
state, chemical bonding and site symmetry. M\"{o}sbauer spectra of Fe-2322
(Fig. 2(a)) shows a single doublet, consistent with a single inequivalent 4c
position of Fe in I4/mmm space group of the unit cell. Isomer shifts $\delta
$ = 0.895(1) mm/s and quadrupole splitting ($\Delta $ = 1.952(1) mm/s) are
well in agreement with the high spin S = 2 of Fe$^{2+}$ in the D$_{2h}$
symmetry site of FeO$_{2}$Se$_{2}$ octahedra.\cite{Fuwa1} However, the $v$ =
0.3 mm/s discrepancy between the experimental and theoretical fits and the
mild asymmetry of peak intensities suggest that another interaction is
superimposed on the main doublet. The Fe$^{2+}$ experiences either an
additional doublet or sextet due to an impurity phase or a high temperature
magnetism in the sample with magnetic dipole interaction of $\mu _{0}H\ll
eV_{zz}Q$ magnitude. However, no known impurity phase, including LaFeO$_{3}$
was able to explain the experimental spectra. In contrast, M\"{o}sbauer
spectra in both La$_{2}$O$_{3}$(Fe$_{0.5}$Mn$_{0.5}$)$_{2}$Se$_{2}$ and La$%
_{2}$O$_{3}$(Fe$_{0.8}$Mn$_{0.2}$)$_{2}$Se$_{2}$ detected Fe-Se phase
impurity with $\delta $ = 0.45(7) mm/s, $\Delta $ = 0.8(1) mm/s. When Mn
substitutes Fe in 1:4 ratio (Fig. 2(b)) there are 8 Fe$^{2+}$ ions in the
first two coordination spheres which contribute to 10 different combinations
after the substitution if all combinations are of equal probability. The
most important contribution to hyperfine interactions is from the first
coordination sphere. With 1:4 Mn to Fe ratio there are 3 different
combinations: with 0, 1 or 2 Mn on four Fe sites. The ratio of these
combinations is 17:20:3 respectively. Distribution of electric quadrupole
splittings centered at 1.756 mm/s, with 0.309 mm/s standard deviation and
pronounced tail near the smaller values confirms uniform distribution of Mn.
The main doublet and minority phase (or interaction) are also present when
one half of Fe$^{2+}$ is replaced by Mn$^{2+}$ in La$_{2}$O$_{3}$(Fe$_{0.5}$%
Mn$_{0.5}$)$_{2}$Se$_{2}$ (Fig. 2(c)). The atomic positions where half of Fe
ions were substituted by Mn are symmetric within the first coordination
sphere around Fe$^{2+}$ that contains FeO$_{2}$Se$_{2}$ and four closest Fe
ions in the Fe-O planes (Fig. 1(a)). This is confirmed by the rather small
widening of the doublet lines ($\Gamma $ = 0.309 (7) mm/s). Decreased values
of isomer shift $\delta $ suggest a change in the charge density on the Fe$%
^{2+}$ and substantial decrease of the octahedra.\cite{Mayer}

\begin{figure}[tbp]
\centerline{\includegraphics[scale=1.1]{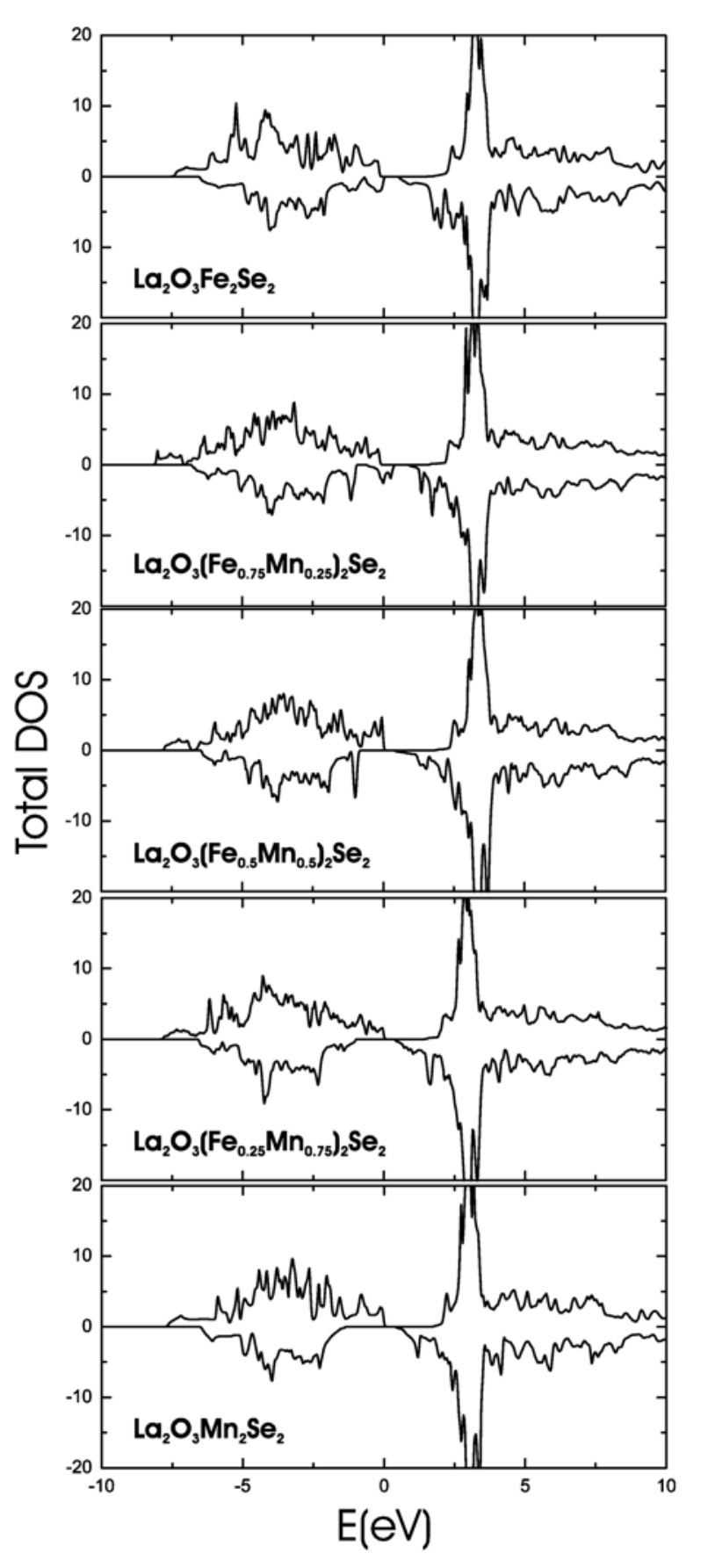}} \vspace*{-0.3cm}
\caption{Total DOS per formula unit for the FM state of (Fe, Mn)-2322. The
Fermi level is set to zero energy. There is a band gap shift of the minority
DOS towards lower energies as Mn substitutes Fe in the lattice.}
\end{figure}

\begin{figure}[tbp]
\centerline{\includegraphics[scale=0.8]{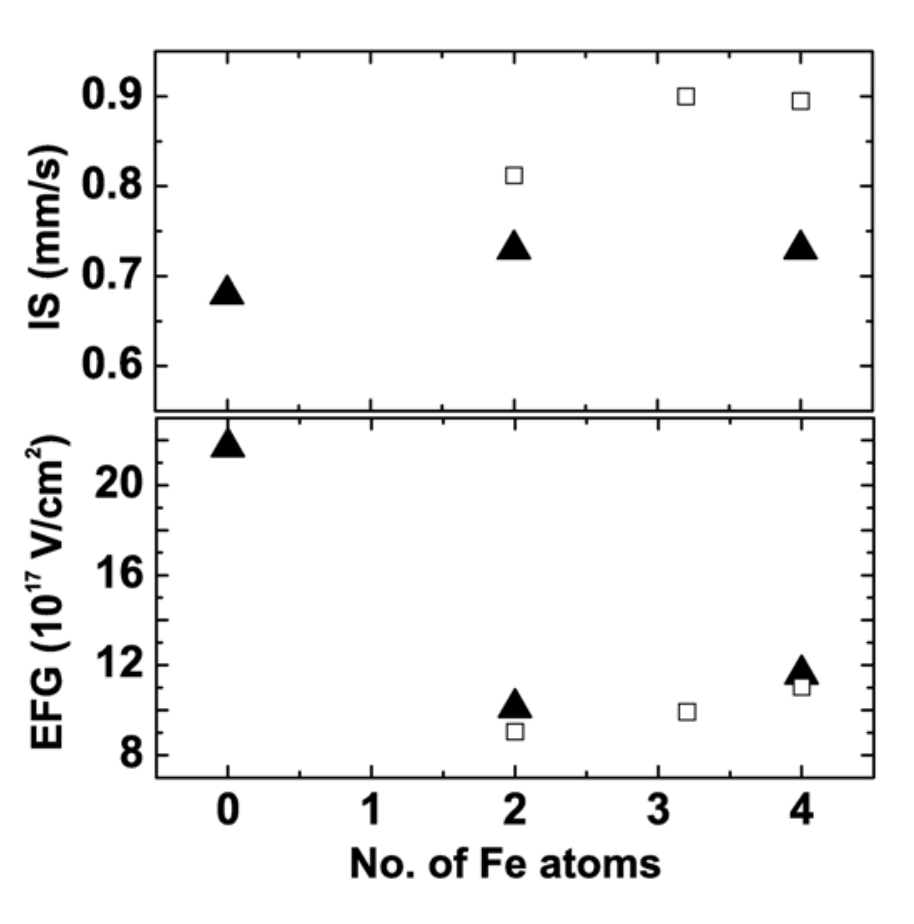}} \vspace*{-0.3cm}
\caption{Calculated (solid triangles) and measured (open squares) EFGs and
ISs plotted as a function of the number of (in-plane) neighboring Fe atoms.
Our 2$\times $2$\times $1 supercells did not allow for geometries with 1 and
3 Fe atoms. The experimental EFGs and ISs are plotted against the average
number of neighboring Fe atoms in La$_{2}$O$_{3}$(Fe$_{0.5}$Mn$_{0.5}$)$_{2}$%
Se$_{2}$, La$_{2}$O$_{3}$(Fe$_{0.8}$Mn$_{0.2}$)$_{2}$Se$_{2}$ and Fe-2322.}
\end{figure}

In order to complement the M\"{o}ssbauer measurements, we performed density
functional theory (DFT) calculations using the WIEN2K all-electron
full-potential APW+lo code.\cite{wien2k} For La, Fe, Mn, Se, and O atoms the
muffin-tin sphere radii were set to 2.38, 1.86, 1.86, 2.5 and 1.62 a.u.. The
basis set functions were expanded up to R$_{mt}$K$_{max}$ = 7. We used the
LDA-Fock-0.25 hybrid functional,\cite{Wu}$^{,}$\cite{Tran} which provides a
viable alternative to the LDA+U approach in describing the electron
correlations in this system. Starting from the Fe-2322 and Mn-2322 pure
structures, we constructed 2$\times $2$\times $1 supercells of La$_{2}$O$%
_{3} $(Fe$_{0.25}$Mn$_{0.75}$)$_{2}$Se$_{2}$, La$_{2}$O$_{3}$(Fe$_{0.5}$Mn$%
_{0.5}$)$_{2}$Se$_{2}$ and La$_{2}$O$_{3}$(Fe$_{0.75}$Mn$_{0.25}$)$_{2}$Se$%
_{2}$ to approximate the mixed Fe-Mn systems. Figure 3 shows the total
density of states (DOS) as calculated for the FM state of the investigated
compounds.

The LDA-Fock-0.25 hybrid functional calculations correctly predict the
Mott-insulating behavior of (Fe, Mn)-2322. The band gaps are similar but
decrease somewhat from pure Fe-2322 (about 0.4 eV) to pure Mn-2322 (about
0.25 eV). The overall shape of the DOS across this group is very similar,
one significant distinction being the band gap shift of the minority DOS
towards lower energies with increased Mn content. In our checkerboard AFM
LDA-Fock-0.25 calculations (carried out only for the end compounds, not
shown here), the band gaps of Fe-2322 and Mn-2322 are found to be 1.85 and
0.99 eV, respectively. Our calculations indicate that the AFM state is more
stable than the FM state by 0.75 eV in Fe-2322 and 0.43 eV in Mn-2322. The
supercell approach also enabled us to calculate the electric field gradient
(EFG) and isomer shift (IS) at Fe situated in several different local
coordinations. In Figure 4 we compare the results of our calculations with
the values extracted from M\"{o}ssbauer spectroscopy. While our calculations
were able to reproduce the values of the measured EFGs, the ISs are found to
be smaller than the experimental values by up to 20 percent. The discrepancy
is because experimental M\"{o}ssbauer includes second order Doppler shift
and chemical isomer shift due to electronic density whereas the theoretical
parts involves only the latter. The calculated EFG of 11.6 $\times $ 10$%
^{21} $ V/m$^{2}$ in pure Fe-2322 agrees very well with the EFG obtained in
similar DFT calculations.\cite{Yayoi}

\begin{figure}[tbp]
\centerline{\includegraphics[scale=0.8]{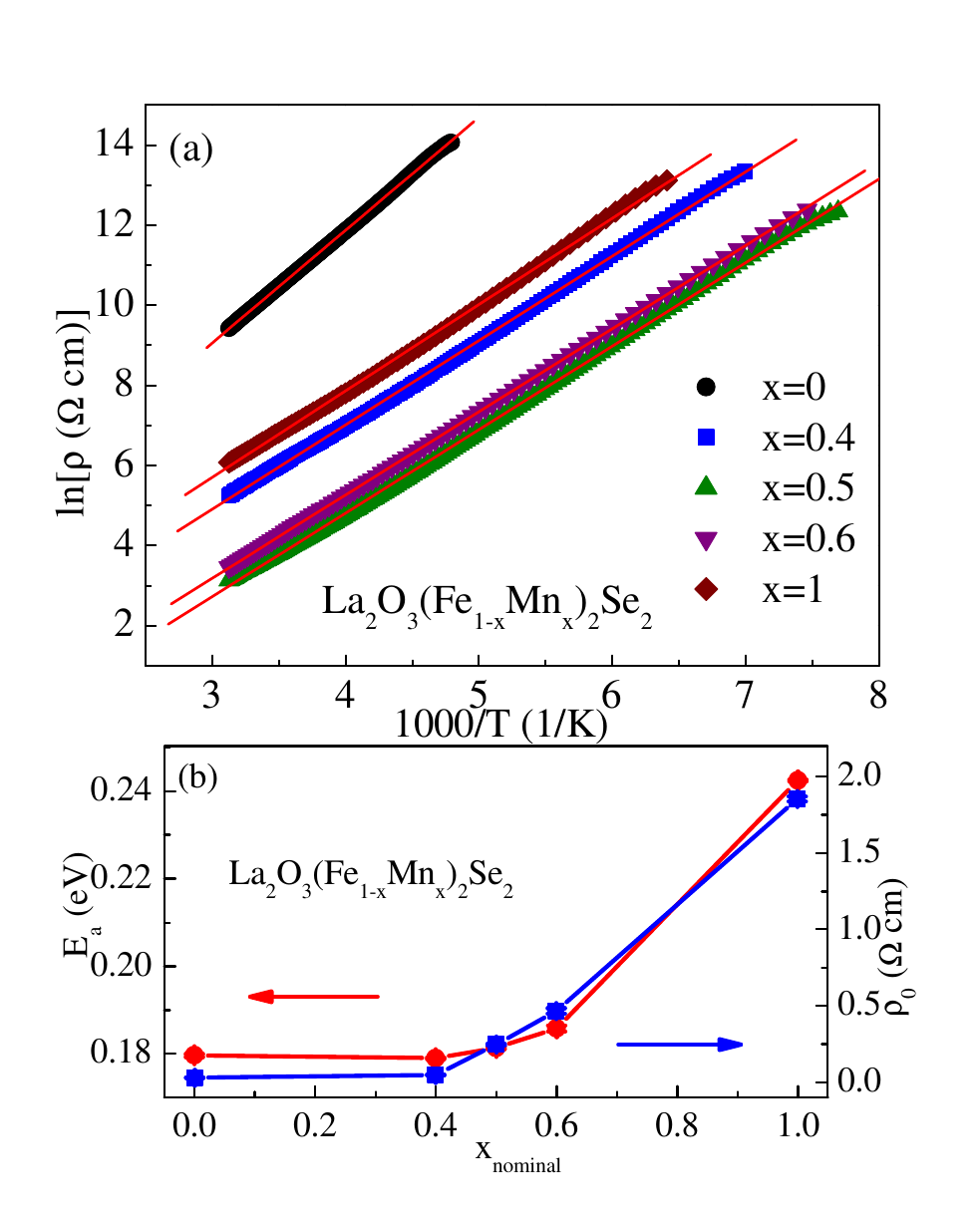}} \vspace*{-0.3cm}
\caption{(a) Temperature dependence of the resistivity for (Fe,Mn)-2322
polycrystals. (b) Activation energies using $\protect\rho =\protect\rho %
_{0}\exp (E_{a}/k_{B}T)$ and the relation between $\protect\rho _{0}$, $%
E_{a} $ and $x_{nominal}$.}
\end{figure}

Neglecting grain boundaries, Mn-2322 resistivity at room temperature is
about 2$\times $10$^{4}$ $\Omega \cdot cm$. This is two orders of magnitude
smaller than Co-2322.\cite{Wang} The resistivities of (Fe,Mn)-2322 exhibit
activated behavior, $\rho =\rho _{0}\exp (E_{a}/k_{B}T)$, where $\rho _{0}$
is the prefactor, $E_{a}$ is the activation energy, $k_{B}$ is the
Boltzmann's constant and $T$ is temperature (Fig. 5(a) and (b)). By fitting
the $\rho (T)$ data from 210 to 320 K we obtain the activation energy $%
E_{a}= $ 0.2425(2) eV for Mn-2322. This is substantially smaller than the
activation energy of Co-2322\cite{Wang} and consistent with the recently
reported value.\cite{Free2} The $E_{a}$ decreases rapidly by $x=$ 0.5
(0.1814(3) eV), then gradually as the Mn/Fe ratio is tuned to Fe-2322
(0.1796(3) eV). The value of $E_{a}$ for Fe-2322 is close to that reported
in the literature.\cite{Zhu} The residual resistivity values ($\rho _{0}$)
exhibit the same trend. These results indicate that the band gap of
(Fe,Mn)-2322 decreases with the contraction of lattice parameters. The
narrowing of band gap is also consistent with sample color changes from
yellowish green in Mn-2322 to black in Fe-2322. Gap values obtained for
Mn-2322 are in agreement with LDA calculations (Fig. 3), however for Fe-2322
and for intermediate alloys there is a disagreement. This suggests increased
importance of strong electronic correlations for increased Fe site occupancy
of Fe/Mn site in (Fe, Mn)-2322 alloys.

\begin{figure}[tbp]
\centerline{\includegraphics[scale=0.7]{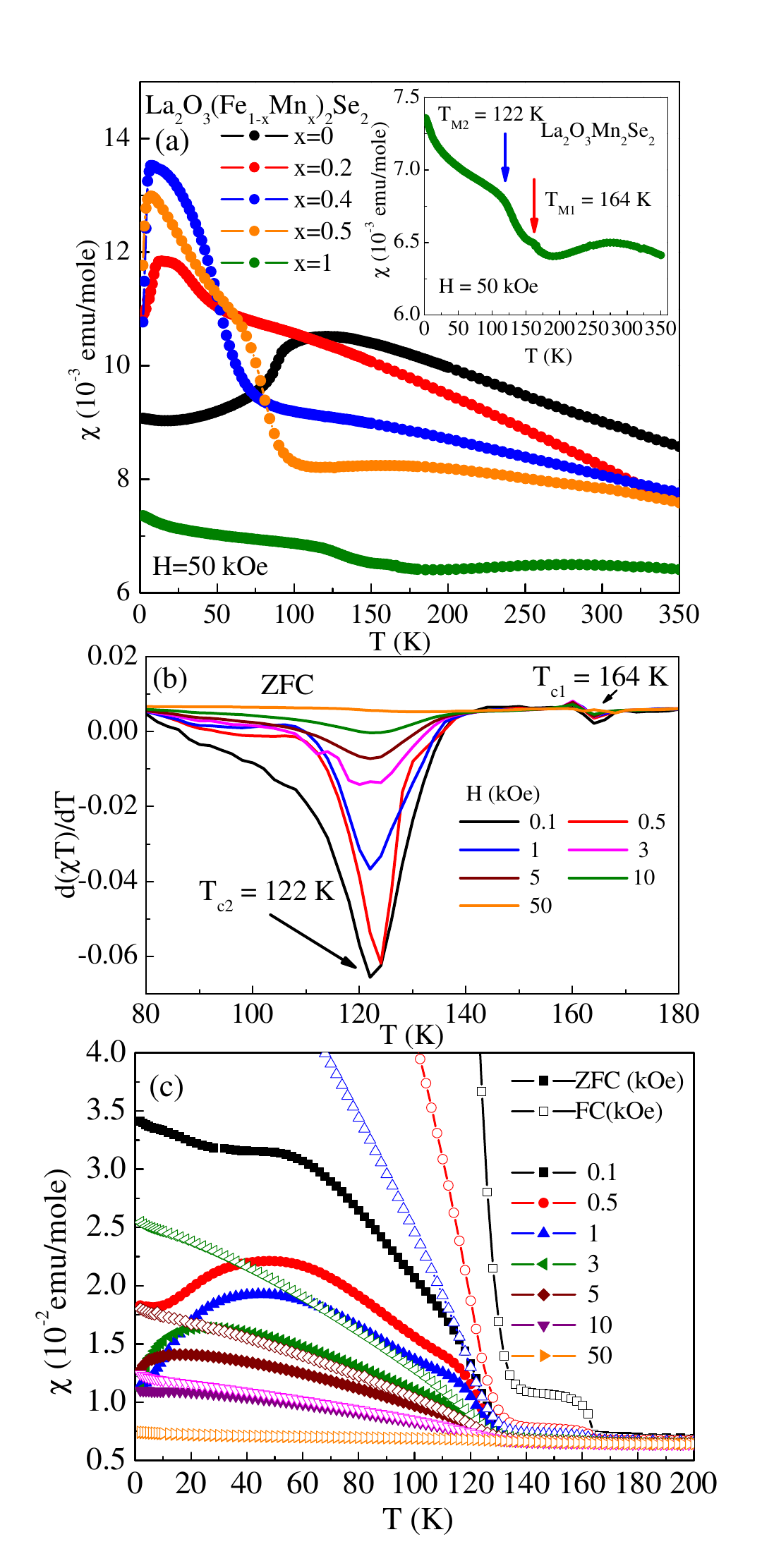}} \vspace*{-0.3cm}
\caption{(a) Temperature dependence of the magnetic susceptibility $\protect%
\chi(T)$ for (Fe,Mn)-2322 taken in 50 kOe. Inset shows enlarged $\protect\chi%
(T)$ for Mn-2322. (b) $d(\protect\chi T)/dT$ of Mn-2322 at various magnetic
fields for ZFC magnetization. (c) ZFC and FC curves of Mn-2322 near magnetic
transitions.}
\end{figure}

Temperature dependence of dc magnetic susceptibility $\chi (T)=M/H$ taken in
50 kOe magnetic field for (Fe,Mn)-2322 with zero field cooling (ZFC) is
shown in Fig. 6(a). For Fe-2322, there is AFM transition at $T_{N}=$ 90 K,
consistent with previous results.\cite{Zhu} With Mn doping, this AFM
transition shifts to lower temperature quickly and disappears in Mn-2322. On
the other hand, there is other FM-like transition emerging and shifting to
higher temperature when Mn enters the lattice. Finally, for Mn-2322, The $%
\chi (T)$ curves of Mn-2322 (inset of Fig. 6(a)) exhibit three features: one
broader maximum, $T_{Max}$ and two anomalies, $T_{M1}$ and $T_{M2}$. $%
T_{Max} $ ($\sim $ 275 K) with no Curie-Weiss behavior up to 350 K implies
that there is a stronger two-dimensional short-range spin correlations in
Mn-2322 than in Fe-2322 at high temperature, reflecting the frustrated
nature of the magnetic structure in Mn-2322.\cite{Ni}$^{,}$\cite{Free2} From
the sharp minimum of the $d(\chi T)/dT$ curves (Fig. 6(b)) we extract $%
T_{M1} $ $\sim $ 164 K, in agreement with previous results.\cite%
{Ni,Liu1,Free2} The $T_{M1}$ anomaly is field independent and therefore
should have considerable ferromagnetic component due to canting and/or
magnetocrystalline anisotropy possibly induced by frustration.\cite{Ni} The
second anomaly (minimum in $d(\chi T)/dT$ at $T_{M2}$ = 122 K (inset of Fig.
6(a))) is also field independent, however it is strongly suppressed in high
fields. This is in agreement with its proposed spin reorientation origin.%
\cite{Ni} The ZFC and field cooling (FC) $\chi (T)$ curves split at $%
T_{M1,2} $ but gradually overlap with magnetic field increase (Fig. 6(c)).
This implies the presence of either magnetic frustration or
magnetocystalline anisotropy. On the other hand, $\chi (T)$ of Mn-2322 show
a hump around $T_{N}$ = 42 K at 1 kOe with ZFC mode (blue curve), similar to
previous results.\cite{Ni}$^{,}$\cite{Free2} It could be due to
superparamagnetic Mn$_{3}$O$_{4}$ nanoparticles contamination below our
diffraction resolution limit.\cite{Free2} Overall, the substantial shift of
the cusp with $x$ and evolution of FM-like increase of magnetization above
the cusp temperature points to development of intrinsic magnetic
interactions from Fe-2322 to Mn-2322. With Mn doping, the part of magnetic
entropy that contributes to long range order rapidly shifts to lower
temperature whereas frustration-induced FM (or canting, anisotropic exchange
interactions) component of magnetization increases\ and high temperature
two-dimensional short range magnetic order becomes stronger as the magnetic
system is tuned toward the lattice with full Mn occupation.

\begin{figure}[tbp]
\centerline{\includegraphics[scale=0.8]{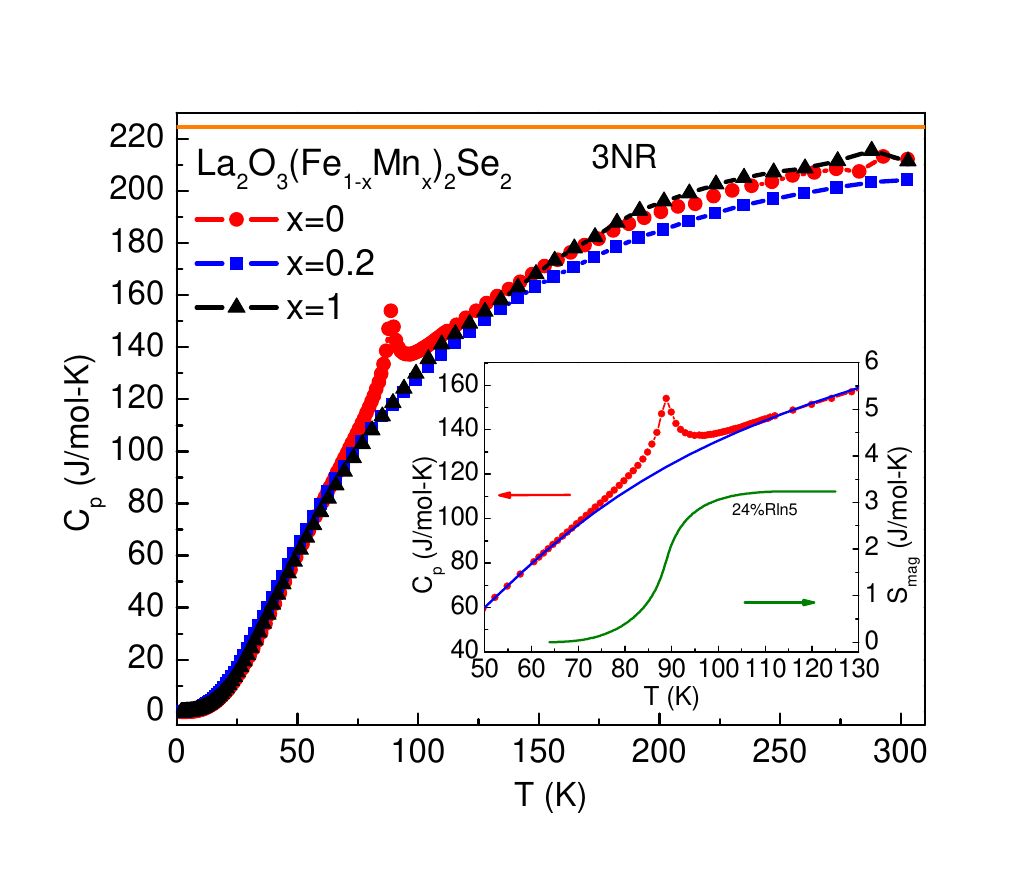}} \vspace*{-0.3cm}
\caption{Temperature dependence of specific heat for (Fe,Mn)-2322
polycrystals. Inset shows the specific heat of Fe-2322 data between 50 K and
130 K. The solid curve represents the phonon contribution fitted by a
polynomial. The right axis and its associated solid curve denote the
magnetic entropy related to the AFM transition of Fe-2322.}
\end{figure}

Specific heat of (Fe,Mn)-2322 (Fig. 7) approaches the value of 3NR at 300 K,
where N is the atomic number in the chemical formula (N = 9) and R is the
gas constant (R = 8.314 J mol$^{-1}$ K$^{-1}$), consistent with the
Dulong-Petit law. On the other hand, Fe-2322 specific heat shows a $\lambda $%
-type anomaly at $T$ = 89 K for Fe-2322 in agreement with the result in the
literature.\cite{Fuwa1} This anomaly is due to a long-range 3D
antiferromagnetic (AFM) ordering of the Fe$^{2+}$. After subtraction of the
phonon contribution ($C_{ph}$) fitted using a polynomial for the total
specific heat, we obtain the magnetic contribution ($C_{mag}$) and calculate
the magnetic entropy using $S_{mag}(T)=\int_{0}^{T}C_{mag}/Tdt$. The derived
$S_{mag}$ is 3.23 J/mol-K at 120 K, which is much smaller than the expected
value ($\sim $ 24\% R$\ln (2S+1)=$ R$\ln 5$) for Fe$^{2+}$ ions with high
spin state (inset of Fig. 7). Note that only about half of that value is
released below $T_{N}$. It suggests that substantial fraction of magnetic
entropy is locked at high temperatures due to possible 2D short-range
magnetic order before long range correlations develop at $T_{N}$.\cite{Ni}
In contrast, there is no $\lambda $-type anomaly for Mn-2322 at the
temperature of magnetic transitions ($T_{M1}=$ 164 K and $T_{M2}=$ 122K),
which is consistent with the previous result and can be ascribed to the
overwhelming release of magnetic entropy due to existence of 2D short-range
order above magnetic transitions.\cite{Ni} The 20\% Mn doping in Fe-2322 has
already resulted in complete release of magnetic entropy above $T_{N}$. If
the entropy is released via the high temperature frustration mechanism, this
suggests that frustration effects are considerably enhanced with Mn doping.
This could explain the absence of Curie-Weiss law in $M(T)$ above $T_{N}$
(Fig. 6(a)).

\section{Conclusion}

La$_{2}$O$_{3}$(Fe$_{1-x}$Mn$_{x}$)$_{2}$Se$_{2}$ compounds are magnetic
with the high spin state for Fe$^{2+}$/Mn$^{2+}$. Intermediate alloys
exhibit long range magnetic order and release magnetic entropy at high
temperature due to possible magnetic frustration. Further neutron scattering
experiments are necessary to provide microscopic information on how
principle competing interactions evolve in this alloy series. FM component
of magnetically ordered ground state and possibly the degree of magnetic
frustration both raise rapidly as Mn enters the lattice. In (Fe,Mn)-2322
compounds long range magnetic order competes with disorder induced by
geometric spin frustration. Atomic disorder on Fe/Mn sites tips this balance
and promotes frustration. Since the balance of all 3 principle interactions $%
J_{1}$-$J_{3}$ can be tuned by Fe/Mn ratio, this system might be mapped to
the 2D pyrochlore where charge order and metal to insulator transition have
been predicted at half filling.\cite{Fujimoto1,Fujimoto2}

\section{Acknowledgements}

The part of this work was carried out at BNL which is operated for the U.S.
Department of Energy by Bookcases Science Associates under grant
DE-Ac02-98CH10886. A portion of this work was benefited from the use of HIPD
at the Lujan Center at Los Alamos Neutron Science Center, funded by DOE
Office of Basic Energy Sciences. Los Alamo National Laboratory is operated
by Los Alamo National Security LLC under DOE Contract No. DE-AC52-06NA25396.
This work has also been supported by the grant No. 45018 from the Serbian
Ministry of Education and Science.


\begin{thebibliography}{99}
\bibitem{Mazin} I. I. Mazin, Nature \textbf{464}, 183 (2010).

\bibitem{Cvetkovic} V. Cvetkovic and Z. Tesanovic, Europhys. Lett \textbf{85}%
, 37002 (2009).

\bibitem{Qazilbash} M. M. Qazilbash, J. J. Hamlin, R. E. Baumbach, L. Zhang,
D. J. Singh, M. B. Maple and D. N. Basov, Nature Phys. \textbf{5}, 647
(2009).

\bibitem{Mayer} J. M. Mayer, L. F. Schneemeyer, T. Siegrist, J. V. Waszczak
and B. Van Dover, Angew. Chem., Int. Ed. Engl. \textbf{31}, 1645 (1992).

\bibitem{Clarke} S. J. Clarke, P. Adamson, S. J. C. Herkelrath, O. J. Rutt,
D. R. Parker, M. J. Pitcher and C. F. Smura, Inorg. Chem. \textbf{47}, 8473
(2008).

\bibitem{Kabbour} H. Kabbour, E. Janod, B. Corraze, M. Danot, C. Lee, M. H.
Whangbo and L. Cario, J. Am. Chem. Soc. \textbf{130}, 8261 (2008).

\bibitem{Wang} C. Wang, M. Q. Tan, C. M. Feng, Z. F. Ma, S. Jiang, Z. A. Xu,
G. H. Cao, K. Matsubayashi and Y. Uwatoko, J. Am. Chem. Soc. \textbf{132},
7069 (2010).

\bibitem{Zhu} J. X. Zhu, R. Yu, H. Wang, L. L. Zhao, M. D. Jones, J. Dai, E.
Abrahams, E. Morosan, M. Fang and Q. Si, Phys. Rev. Lett. \textbf{104},
216405 (2010).

\bibitem{Free1} D. G. Free and J. S. O. Evans, Phys. Rev. B \textbf{81},
214433 (2010).

\bibitem{Wu} H. Wu, Phys. Rev. B \textbf{82}, 020410 (2010).

\bibitem{Fuwa1} Y. Fuwa, M. Wakeshima and Y. Hinatsu, J. Phys. Cond. Matt.
\textbf{22}, 346003 (2010).

\bibitem{Fuwa2} Y. Fuwa, T. Endo, M. Wakeshima, Y. Hinatsu and K. Ohoyama,
J. Am. Chem. Soc. \textbf{132}, 18020 (2010).

\bibitem{McCabe} E. E. McCabe, D. G. Free, B. G. Mendis, J. S. Higgins and
J. S. O. Evans, Chem. Mater. \textbf{22}, 6171 (2010).

\bibitem{Ni} N. Ni, E. Climent-Pascual, S. Jia, Q. Huang and R. J. Cava,
Phys. Rev. B \textbf{82}, 214419 (2010).

\bibitem{Liu1} R. H. Liu, J. S. Zhang, P. Cheng, X. G. Luo, J. J. Ying, Y.
J. Yan, M. Zhang, A. F. Wang, Z. J. Xiang, G. J. Ye and X. H. Chen, Phys.
Rev. B \textbf{83}, 174450 (2011).

\bibitem{Free2} D. G. Free, N. D. Withers, P. J. Hickey and J. O. Evans,
Chem. Mater. \textbf{23}, 1625 (2011).

\bibitem{Ling} C. D. Ling, J. E. Millburn, J. F. Mitchell, D. N. Argyriou,
J. Linton and H. N. Bordallo, Phys. Rev. B \textbf{62}, 15096 (2000).

\bibitem{Low} U. L\"{o}w, V. J. Emery, K. Fabricius and S. A. Kivelson,
Phys. Rev. Lett. \textbf{72}, 1918 (1994).

\bibitem{Yildirim} T. Yildirim, Phys. Rev. Lett. \textbf{101}, 057010 (2008).

\bibitem{Si} Q. Si and E. Abrahams, Phys. Rev. Lett. \textbf{101}, 076401
(2008).

\bibitem{Han} M. J. Han, Q. Yin, W. E. Pickett and S. Y. Savrasov, Phys.
Rev. Lett. \textbf{102}, 107003 (2009).

\bibitem{Li} S. Li, C de la Cruz, Q. Huang, Y. Chen, J. W. Lynn, J. Hu, Y.
L. Huang, F. C. Hsu, K. W.Yeh, M. K. Wu and P. C. Dai, Phys. Rev. B \textbf{%
79}, 054503 (2009).

\bibitem{Hunter} B. Hunter, Int. Un. of Cryst. Comm. Newsletter \textbf{20}
(1998).

\bibitem{Larson} A. C. Larson, and R. B. Von Dreele, General structure
analysis system, Report No. LAUR-86-748, Los Alamos National Laboratory, Los
Alamos, NM, (2000).

\bibitem{Toby} B. H. Toby, J. Appl. Crystallogr. \textbf{34} 201 (2001).

\bibitem{Brand} R. A. Brand, WinNormos M\"{o}ssbauer fitting program,
Universit\"{a}t Duisburg 2008

\bibitem{Brown} I. D. Brown and D. Altermatt, Acta Crystalogr. Sect. B
\textbf{41}, 244 (1985).

\bibitem{Brese} N. E. Brese and M. O'Keefe, Acta Crystalogr. Sect. B \textbf{%
47}, 192 (1991).

\bibitem{wien2k} P. Blaha, K. Schwarz, G. Madsen, D. Kvasnicka, and J.
Luitz, "Wien2k, an augmented plane wave + local orbitals program for
calculating crystal properties,", (Karlheinz Schwarz, Techn. Universi\"{a}t
Wien, Austria), 2001. ISBN 3-9501031-1-2.

\bibitem{Tran} F. Tran, P. Blaha, K. Schwarz, and P. Nov\'{a}k, Phys. Rev. B
\textbf{74}, 155108 (2006).

\bibitem{Yayoi} F. Yayoi, W. Makoto, and H. Yukio, J. Phys.: Condens. Matter
\textbf{22}, 346003 (2010).

\bibitem{Fujimoto1} S. Fujimoto, Phys. Rev. Lett. \textbf{89}, 226402 (2002).

\bibitem{Fujimoto2} S. Fujimoto, Phys. Rev. B \textbf{67}, 235102 (2003).
\end{thebibliography}
\end{document}